# Diversity, Representation, and Accessibility Concerns in Game Development

Nooshin Darvishinia (1), Todd Goodson (2)

(1) Doctoral Student, Department of Curriculum and Instruction, Kansas State University, USA.

Email: nooshindvn@ksu.edu

(2) Chair of the Department of Curriculum and Instruction, Associate Dean of Teacher Education and Accreditation, Kansas State University, USA.

Email: tgoodson@ksu.edu

**Abstract**

This study delves into the key issues of representation and accessibility in game development. Despite their societal significance, video games face ongoing criticism for lacking diversity in both the workforce and content, excluding marginalized gamers. This study explores game-based learning (GBL) while emphasizing the importance of accurate representation, particularly in educational settings to enhance engagement and learning outcomes. Our research findings revolve around the perspectives of a professional in the gaming industry and the challenges associated with creating accessible games. By providing actionable insights, it aims to influence regulatory reforms, industry practices, and game creation itself, to foster diversity, representation, and accessibility in the video game industry. In doing so, we seek to promote a more inclusive and equitable future in the educational gaming world.

> Keywords: Game-based learning (GBL); representation; accessibility; gaming industry

**Introduction**

In game design, representation refers to the accurate and inclusive display of a variety of individuals, cultural experiences, and perspectives, particularly in an educational environment. Representation guarantees that different individuals and groups may see themselves reflected in the game's content. This could lead to a significant impact on participation, learning outcomes, and society's awareness of diversity and inclusion. Sousa, Neves, and Damásio, (2022) discuss representation in the context of people with disabilities, notably those with intellectual impairments. These researchers believe that it is critical to investigate not just concerns regarding representation but also questions of representativeness in the integration of individuals with disabilities, including those with intellectual impairments, both in academics and in the field of gaming.

While video games have firmly established their position in society, they have also come under significant criticism for the lack of diversity in both the industry's workforce and the representation of diverse characters within games. According to a study conducted by Jeremic (2023), faculty members pointed out that the gaming industry and game education are predominantly male-dominated, resulting in the underrepresentation of women. The underlying issue appears to be the industry's tendency to perceive games as purely technical artifacts rather than cultural ones, which, in turn, limits their influence on society. Furthermore, the industry has faced criticism for its approach to accessibility, which often marginalizes gamers with diverse abilities and experiences. These exclusions limit a game's potential audience and overlook a broad range of experiences and perspectives that could enrich the gaming context. Game-based learning (GBL) emerged from gaming research in the mid-1950s, and researchers began researching and practicing integrating games into education in the 1980s. Game-based learning is characterized as a sort of gameplay with specified learning outcomes (Shaffer, Halverson, Squire, & Gee, 2005). Serious games, analog educational games (Lickteig, 2017), and digital educational games are the three concepts that are always utilized in GBL studies (Pan et al., 2021).

According to the study conducted by Tlili et al. (2021) representations in GBL are vital for learners with disabilities. The study emphasizes the efficacy of game-based learning in improving cognitive skills, social connections, and academic achievement for these students. However, for such learning to be truly effective, it must be adapted to the specific demands of various disadvantaged groups. The study underlines the need to actively incorporate learners with disabilities in the design and implementation phases in order to ensure that their opinions and experiences are accurately represented. Collaboration among numerous stakeholders, such as special education instructors, parents, and specialists from other sectors, is also significant. This broad approach guarantees that game-based learning is not just accessible but also adaptable, meeting the unique needs of learners with disabilities.

There are various challenges associated with educational games. One of these challenges is that these games are either too appealing but fail to represent learning objectives, or they are too instructional but fail to capture learners' attention (Zhang and Liu, 2007). Another study (Plass, Homer, & Kinzer, 2015) discusses the difficulties in designing and implementing successful game-based learning experiences. They underline the significance of more studies to better understand the influence of games on learning outcomes and to provide guidelines for incorporating games into educational practice.

*Accessibility in Games*

Games have been proven to enhance student motivation and engagement and can be associated with behavioral change (de Freitas, 2018); however, it appears that the significance of accessibility features has been ignored. While game designers and developers should consider incorporating a variety of customizable settings in their products, some may not take these features into account.

Socioeconomic variables pose significant obstacles to game-based learning (e.g., Arztmann, Hornstra, Jeuring, & Kester, 2023) particularly among learners who may suffer a variety of financial and resource constraints. The digital divide exacerbates this issue, as learners from lower socioeconomic backgrounds may lack reliable internet access or the devices needed to engage with game-based learning platforms effectively. Even when digital resources are available, they may be shared among multiple family members, making it difficult for adult learners to allocate sufficient time for game-based learning activities.

To solve this, game developers should consider low-bandwidth and offline versions, guaranteeing that learners may participate even without a high-speed internet connection. Similarly, sliding-scale price structures or freemium versions might make these learning tools more affordable. It is critical for game-based learning platforms to offer a wide range of hardware specifications, from high-end PCs and tablets to more affordable ones, to ensure inclusion for learners from all socioeconomic backgrounds.

The absence of visual accommodations like high-contrast viewing options limits participation for individuals with visual impairments (von Gillern & Nash, 2023). Game designers and developers should think about including a wide range of configurable graphic options in their products. Recommendations such as font size adjustments, screen magnifiers, alternative information conveyance methods, high contrast modes or luminosity alterations (Torrente et al., 2014), flexible color schemes, and icon scaling appeal to varied visual demands. Furthermore, extensive testing should be performed with feedback from users with visual difficulties to verify that these functionalities satisfy their unique needs.

The lack of auditory accommodations (Beeston, Power, Cairns, & Barlet, 2018) such as subtitles, restricts access for those with hearing impairments. All game-based learning platforms should have robust support for subtitles, text prompts, and even text-to-speech features. In addition to meeting accessibility guidelines, designers should solicit feedback from the hearing-impaired community to ensure that auditory features are effectively implemented. This can help confirm that all auditory information is clearly and effectively communicated through text.

*Research questions*

By giving voice to an industry insider's experiences and viewpoints, we aim to find actionable insights that may influence regulatory reforms, industry practices, and game creation itself. This study aims to give essential insights that will enable game developers, and stakeholders to design games that will appeal to a wider audience and promote inclusion. Additionally, the success of this study will benefit not merely the game development sector, but it will also enhance the gaming experience for gamers of various backgrounds and skills. It promotes diversity, representation, and accessibility in every aspect of media and entertainment, encouraging a more inclusive and equitable future in the video game industry. In pursuit of a more inclusive and equitable future in the video game industry, this study sets out to answer the following critical research questions:

How do professionals view diversity and representation in game development?

What challenges exist in creating accessible games, and how can they be overcome?

*Definition of key terms*

Professionals: In the context of this study, "professionals" refer to individuals who are actively engaged in the fields of game development, which includes game designers, developers, marketers, and other industry experts.

Diversity: Diversity, within the scope of this research, encompasses various aspects, including race, gender, socioeconomic class and disabilities. It focuses on acknowledging and embracing differences in these areas to create a more inclusive environment in the gaming industry.

Representation: In the context of this study, "representation" pertains to the portrayal of characters that are developed within video games. It involves ensuring that game characters authentically and inclusively reflect a wide range of individuals, cultural experiences, and perspectives, especially within educational settings.

Game Development: "Game development" encompasses the entire process involved in creating and bringing video games to the market. This process includes the conceptualization, design, development, and marketing of games, as well as the associated industry practices.

Challenges: The "challenges" discussed in this study refer to deviations from established gaming traditions that can provoke resistance from within the gaming industry. These challenges may include issues related to diversity, representation, and accessibility, and they can hinder the progress of the industry.

Accessible Games: "Accessible games" are those designed to ensure that individuals with diverse abilities and experiences can fully engage with and enjoy the gaming experience. This includes accommodating users with physical, sensory, or cognitive impairments, and addressing potential obstacles to their participation in video games.

**Theoretical Foundation**

Definitions of game-based learning mostly emphasize that it is a type of gameplay with defined learning outcomes (Shaffer, Halverson, Squire, & Gee, 2005). Constructivism is an educational paradigm that asserts that learners are active constructors of their comprehension via experiences and interactions rather than passive absorbers of knowledge (e.g., Peters-Richardson, 2023; Mula Tshilumba Nestor, 2022). The learner is encouraged to participate in problem-solving and critical thinking in this constructivist framework to integrate new material into existing cognitive systems. Digital game-based learning (DGBL) is a contemporary illustration of constructivist ideas in action, providing an interactive environment with a focus on cognitive and social engagement. DGBL delivers a personalized and interesting educational experience with features such as problem-solving tasks, immediate evaluations, and multiplayer options.

These digital features are consistent with Vygotsky's concept of the "zone of proximal development" (1978), which emphasizes the necessity of socially mediated, personalized learning experiences. Vygotsky established this notion in the context of child development. Adaptive difficulty levels and fast feedback mechanisms increase engagement and motivation, while score systems, leaderboards, and achievements tap into both inner and extrinsic motivating aspects. The realism of replicated real-world circumstances makes the learning setting more relevant and accessible, promoting quicker absorption and real-world application of knowledge.

Additionally, the interactive and iterative design of digital games fosters metacognitive skills through strategies such as visualization, self-monitoring, and summarization (Hossu & Roman, 2023), allowing learners to reflect on their cognitive processes and, in turn, become more effective learners.

In today's digital age, game-based learning is a powerful platform that dynamically integrates constructivist and experiential learning concepts, all while leveraging technology capabilities. The virtual worlds inside these digital games, in particular, create rich, immersive settings that not only allow learners to construct their own knowledge (compatible with constructivist theories) but also provide them with hands-on experiences that are essential to experiential learning (Kolb, 2014). This dynamic is reinforced further by Lickteig's (2017) findings, which demonstrate that games engage learners in a variety of cognitive situations, both individually and socially.

One of the key technological advantages of game-based learning is the capacity to offer real-time data and feedback. This feature allows games to adjust to each learner's present zone of proximal development, operating as an adaptive learning environment. As a result, the game dynamically adjusts the amount of challenge and assistance, resulting in a genuinely personalized learning experience that aids learners in creating their knowledge. Together, these features demonstrate how game-based learning, with its unique characteristics, enhances both constructivist and experiential learning paradigms.

**Methodology**

*Sampling and Participants*

For this study, a single-case qualitative research design <u>was</u> employed, focusing on an in-depth interview with a professional experienced in software engineering and game development. The participant was purposefully selected based on their expertise, ensuring that their insights provided valuable perspectives on diversity, representation, and accessibility in game development. The participant was contacted via email to seek consent for participation in the research project. The email provided detailed information about the research objectives, the topics to be covered during the interview, and assurances of confidentiality and anonymity. Formal consent was obtained from the participant. The individual is an experienced software developer with an extensive background in game development who provided rich insights into the realm of gaming, game design, and representation.

*Data Collection*

The primary data collection method was a one-on-one Zoom interview with the selected participant. The interview was scheduled at a mutually convenient date and time. Before the interview, a set of pre-determined questions, aligned with the research themes of diversity, representation, and accessibility in game development, were prepared. (Interview questions can be found in **Appendix A)**. The questions were designed to elicit comprehensive insights into the participant's experiences and perspectives. The interview session lasted approximately 60 minutes and was audio-recorded with explicit permission from the participant. (Interview notes can be found in **Appendix B**). The audio recording served as the primary source of data for the study. The audio recording was transcribed using

Microsoft OneNote's transcription services to expedite the data processing phase. The automatically generated transcription was meticulously reviewed alongside the original audio recording to ensure accuracy and reliability. Any identifying information within the transcript was carefully anonymized to uphold the participant's confidentiality and privacy. The transcribed data underwent a rigorous qualitative review using thematic analysis. This process involved identifying patterns, themes, and insights related to the research objectives. The analysis was conducted systematically to extract meaningful information from the participant's responses.

    In the process of participant selection, specific criteria were employed to ensure the selected participant possessed a diverse range of experiences in software engineering and game development, contributing valuable insights to the study. While the selection aimed at capturing a nuanced understanding of diversity, representation, and accessibility in game development, it is acknowledged that inherent biases may exist in the choice of a single participant. To mitigate potential biases, efforts were made to select a participant with a broad perspective, and any limitations associated with this selection approach were duly acknowledged. Ethical considerations extended beyond obtaining formal consent. Confidentiality and privacy were prioritized throughout the research process, with identifying information meticulously anonymized in the transcription phase. The interview questions, provided in Appendix A, were carefully developed and aligned with research themes, and a pilot test was conducted to enhance clarity and relevance. Reflexivity was acknowledged, with the researcher's potential biases recognized and actively managed throughout data collection and analysis. The choice of Microsoft OneNote for transcription was made for its efficiency, and the accuracy of transcriptions was validated through meticulous review alongside the original audio recording.

*Data Analysis*

Qualitative data for this research was collected through in-depth, semi-structured interviews, employing the thematic analysis method to extract meaningful insights from the participant's responses. A thorough review of the interviews identified fifteen codes that encapsulate various facets of the interviewee's perspectives on diversity, representation, and accessibility in game development.

    The first theme, "Professional Background and Focus," emerged from codes related to the interviewee's professional trajectory, including 'Not focused on Game Development,' '15 Years in the Industry,' 'Diverse Product Experience,' and 'Limited VR Development Experience.' Subcodes such as 'Software Engineering Background,' 'Current Role,' 'Coding Experience,' and 'Years in Industry' were drawn, revealing the participant's substantial experience and authority in software engineering, adding credibility to their opinions.

The second theme, "Interest and Perspectives on Emerging Technologies," focused on the code 'AR Development Excitement.' Subcodes like 'Universal Coding Skills,' 'Interest in AR,' and 'Tech Trend Awareness' illustrated the interviewee's broad understanding of coding skills and their particular enthusiasm for emerging technologies like augmented reality (AR).

The third theme, "Approach to Game and App Design," encompassed codes related to the interviewee's involvement in educational apps and their description of the design process. Subcodes, including 'Psychology in Games,' 'Design Importance,' and 'Importance of User Feedback,' highlighted the interviewee's awareness of psychological factors contributing to a game's success and their emphasis on design and user feedback.

    The fourth theme, "Perspectives on Representation and Diversity," covered codes related to the interviewee's perception of underrepresentation and thoughts on diversity in character and

game design, including 'Internet Connectivity Issues.' Subcodes such as 'Understanding of Underrepresentation,' 'Industry Gender Imbalance,' 'Lack of Diversity in Games,' and 'User Testing Demographics' demonstrated the interviewee's acknowledgment of social issues in the industry, suggesting areas for improvement.

The fifth theme, "Considerations in Game Development," included codes related to the interviewee's understanding of the target demographic identification process and concerns about accessibility. Subcodes like 'Market Research Methods,' 'Targeted Marketing Strategies,' and 'Not Involved in Initial Planning' reflected the interviewee's comprehension of industry techniques and the importance of targeted campaigns.

The sixth theme, "Importance of User Feedback," highlighted the code 'Emphasis on User Feedback,' though no specific subcodes were mentioned under this theme.

The seventh theme, "Challenges and Solutions in Game Accessibility," encompassed codes related to concerns about accessibility, internet connectivity issues, and opinions on updates related to representation or accessibility. Subcodes, including 'Accessibility Planning' and 'Internet Accessibility as a Human Rights Issue,' indicated that the interviewee viewed accessibility not just as a game development concern but as a broader human rights issue.

The eighth theme, "Balancing Business and Ethical Concerns," covered codes related to opinions on updates related to representation or accessibility and balancing equity with sales and marketing. Subcodes like 'Prototyping Challenges' and 'Low Priority for Representation Updates' showcased the interviewee's awareness of the practical challenges involved in balancing ethical considerations with business priorities.

**Results**

Upon analyzing the collected qualitative data and identifying codes and themes through thematic analysis, two prominent themes emerged as central to the participant's perspectives on diversity, representation, and accessibility in game development.

*Theme 1: Perspectives on Representation and Diversity*

This theme encompasses the interviewee's nuanced views on representation and diversity in the gaming industry. The following codes and subcodes were recurrent in the participant's responses:

*Codes*

- Underrepresentation Perception
- Thoughts on Diversity in Character and Game Design
- Internet Connectivity Issues

*Subcodes*

- Understanding of Underrepresentation
- Industry Gender Imbalance
- Lack of Diversity in Games
- User Testing Demographics

Figure 1: Perspectives on Representation and Diversity

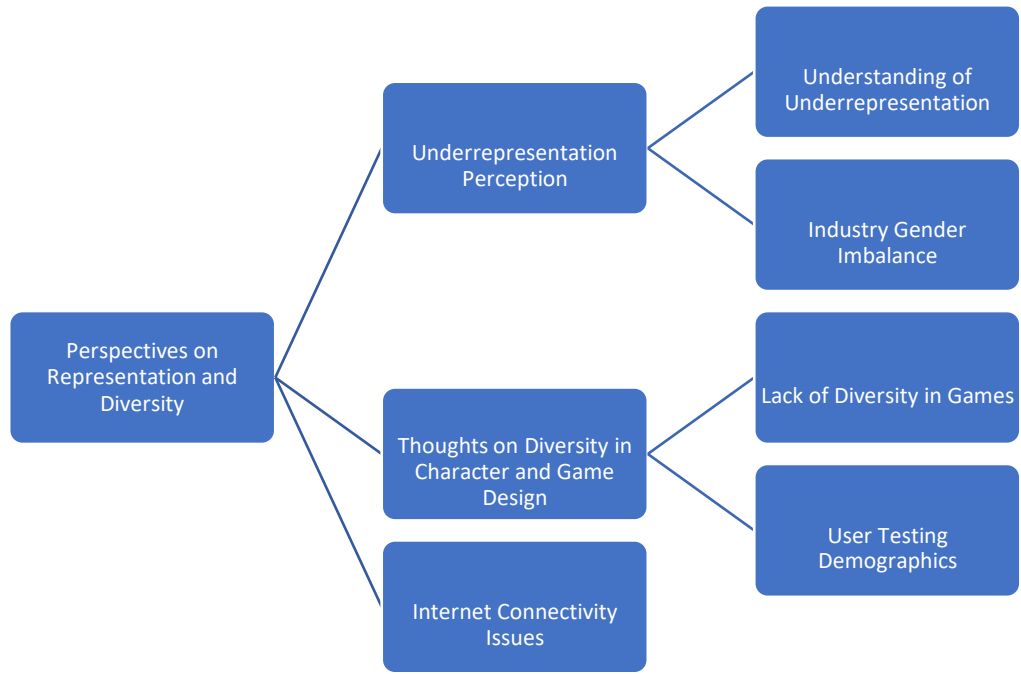

Figure 1 depicts the organization of codes and subcodes under the theme of "Perspectives on Representation and Diversity." The primary theme is subdivided into three main codes, each with specific subcodes that provide a detailed exploration of the participant's insights.

The participant expressed awareness of the industry's underrepresentation issues, particularly emphasizing the gender imbalance and the lack of diversity in games. Insights into user testing demographics shed light on the participant's acknowledgment of the need for diverse perspectives during the testing phase. Additionally, the mention of internet connectivity issues suggests a broader consideration of accessibility concerns, intertwining the themes of representation and accessibility.

*Theme 2: Challenges and Solutions in Game Accessibility*

This theme revolves around the participant's concerns, opinions, and potential solutions related to game accessibility. The following codes and subcodes emerged consistently:

*Codes*

- Concerns about Accessibility
- Internet Connectivity Issues
- Opinions on Updates Related to Representation or Accessibility

*Subcodes*

- Accessibility Planning
- Internet Accessibility as a Human Rights Issue

Figure 2. Structure of Codes and Subcodes Related to "Challenges and Solutions in Game Accessibility "

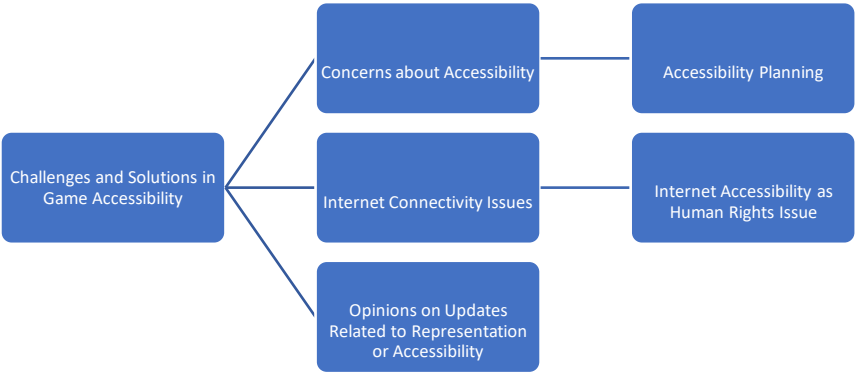

Figure 2 illustrates the structure of codes and subcodes related to "Challenges and Solutions in Game Accessibility." The theme encompasses concerns, opinions, and potential solutions, further subdivided into specific codes and subcodes that capture the participant's perspectives on accessibility in game development.

The interviewee expressed notable concerns about accessibility in game development, emphasizing the planning required to ensure inclusivity. The participant went beyond the conventional scope of game development, viewing internet accessibility as a human rights issue. Opinions on updates related to accessibility underscored the participant's recognition of the ongoing nature of efforts to improve representation and accessibility in the gaming industry.

These two central themes, "Perspectives on Representation and Diversity" and "Challenges and Solutions in Game Accessibility," provide a comprehensive understanding of the participant's insights and contribute to the broader discourse on fostering inclusivity and diversity in game development.

**Conclusion**

This study not only reveals the diverse views of an industry insider, but it also has practical implications that point to a transformational path in the educational game production scene. The investigation of representation reflects a thorough knowledge of the industry's difficulties, specifically gender inequities and a lack of diversity in game content. Placing these problems into the larger framework of user testing demographics and internet connection emphasizes the necessity for a thorough and inclusive approach, particularly in educational contexts.

The emphasis on accessibility challenges, along with a strong focus on diversity planning, extends beyond the conventional parameters of game production, into the domain of educational effect. Recognizing online accessibility as a human rights problem displays a greater understanding of the societal implications and duties involved with gaming accessibility, particularly in the context of instructional games.

This collection of findings illuminates the relationship between representation, diversity, and accessibility in shaping the future of educational game development. As games become increasingly integral to educational narratives, incorporating multiple perspectives and ensuring accessibility becomes not just a gaming industry imperative but a crucial facet of creating enriched learning platforms. This research, therefore, aspires to be a catalyst for positive transformation, advocating for diversity, authentic representation, and accessibility in all forms of media and entertainment, particularly those utilized for educational purposes.

It is critical to move the industry toward embracing diversity in educational material and prioritizing accessibility in design when navigating the constantly changing environment of educational gaming. We are steering towards a future where educational gaming serves as an example of inclusiveness by actively participating in continuing efforts to establish an educational gaming environment that resonates with a diverse student population. This vision aims to enable learners of all backgrounds and abilities to participate meaningfully in the fascinating and immersive worlds that the educational gaming industry has to offer.

**References**

Arztmann, M., Hornstra, L., Jeuring, J., & Kester, L. (2023). Effects of games in STEM education: a meta-analysis on the moderating role of student background characteristics. Studies in Science Education, 59(1), 109-145. https://doi.org/10.1080/03057267.2022.2057732

Beeston, J., Power, C., Cairns, P., & Barlet, M. (2018). Accessible Player Experiences (APX): The Players. In K. Miesenberger & G. Kouroupetroglou (Eds.), Computers Helping People with Special Needs (Vol. 123, LNCS 10896, pp. 245-253). *Springer*. https://www-users.york.ac.uk/~pc530/pubs/Beeston_ICCHP2018.pdf

de Freitas, S. (2018). Are Games Effective Learning Tools? A Review of Educational Games. *Journal of Educational Technology & Society*, *21*(2), 74–84. http://www.jstor.org/stable/26388380

Hossu, R., & Roman, A. F. (2023). Teaching Metacognitive Reading Strategies to Primary School Students Through Digital Games. In I. Albulescu, & C. Stan (Eds.), Education, Reflection, Development - ERD 2022, vol 6. European Proceedings of Educational Sciences (pp. 677-685). European Publisher. https://doi.org/10.15405/epes.23056.62

Jeremic, J. (2023). Qualitative Analysis of the Position of Social Issues in Video Game Education (Doctoral dissertation, Simon Fraser University). Retrieved from https://summit.sfu.ca/_flysystem/fedora/2023-05/etd22385.pdf

Kolb, D. A. (2015). *Experiential Learning: Experience as the Source of Learning and Development* (2nd ed.). Pearson Education. ISBN: 978-0-13-389240-6

Lickteig, S. J. (2020). *Metagaming: Cognition in gaming environments and systems* (Doctoral dissertation, Kansas State University). Retrieved from https://krex.k-state.edu/bitstream/handle/2097/40814/SethLickteig2020.pdf?sequence=1

Mula Tshilumba Nestor. (2022). PROMOTING COLLABORATIVE LEARNING TO FACILITATE COMMUNICATION AMONG CONGOLESE EFL LEARNERS. *Galaxy International Interdisciplinary Research Journal*, *10*(4), 595–606. Retrieved from https://giirj.com/index.php/giirj/article/view/2334

Pan, L., Tlili, A., Li, J., Jiang, F., Shi, G., Yu, H., & Yang, J. (2021). How to Implement Game-Based Learning in a Smart Classroom? A Model Based on a Systematic Literature Review and Delphi Method. Frontiers in Psychology, 12(749837). https://doi.org/10.3389/fpsyg.2021.749837

Peters-Richardson, J. Y. (2023). An analysis of teachers' experiences with professional development and developing learners' 21st-century skills (Doctoral dissertation, Unicaf University). https://cdn.unicaf.org/websites/unicaf/wp-content/uploads/2023/09/Jacqueline-Peters-Richardson-FINAL-Thesis.pdf

Plass, J. L., Homer, B. D., & Kinzer, C. K. (2015). Foundations of game-based learning. *Educational Psychologist*, 50(4), 258–283. https://doi.org/10.1080/00461520.2015.1122533

Shaffer, D. W., Squire, K. R., Halverson, R., & Gee, J. P. (2005). *Video Games and the Future of Learning* (WCER Working Paper No. 2005-4). Wisconsin Center for Education Research, University of Wisconsin–Madison. https://files.eric.ed.gov/fulltext/ED497016.pdf

Sousa, C., Neves, J. C., & Damásio, M. J. (2022). The Pedagogical Value of Creating Accessible Games: A Case Study with Higher Education Students. *Multimodal Technologies and Interaction*, 6(2), 10. https://doi.org/10.3390/mti6020010

Torrente, J., del Blanco, Á., Serrano-Laguna, Á., Vallejo-Pinto, J. Á., Moreno-Ger, P., & Fernández-Manjón, B. (2014). Towards a Low-Cost Adaptation of Educational Games for People with Disabilities. ***Computer Science and Information Systems, 11***(1), 369–391. **https://doi.org/10.2298/CSIS121209013T**

von Gillern, S., & Nash, B. (2023). Accessibility in video gaming: An overview and implications for English language arts education. *Journal of Adolescent & Adult Literacy*, 66(6), 382–390. https://doi.org/10.1002/jaal.1284

Zhang, W., & Liu, J. (2007). The Examined Thought about Essence and Value of Edugame—Looking at Combination of Education and Game From Game Angle [J]. *Open Educ. Res*, *5*, 64-68.

**Disclosure statement**

There is no conflict of interest involved in this study.

**Funding**

We did not receive funding for this study.

# Appendix A

## Interview Questions

**Background**

**How long have you been in gaming?**

What kinds of products have you been involved with? Specific projects he has been involved with?

**What does "Underrepresented" mean?**

How do you define underrepresented groups? What do you consider as presented?
To what extent do you consider representations in developing visual elements?

**Conceptualization and Planning**

How do you identify target demographics during the initial planning stages? Are underrepresented groups specifically considered? What steps are taken at this stage to ensure that the educational content will be accessible to a diverse audience?

**Design**

Do you consider the representation of different characters?

Are the game's visual elements (like characters, settings, and items) representative of diverse cultures, genders, and abilities?

Are there specific design elements included to make educational content culturally relevant and relatable to a diverse player base?

**Prototyping**

Do you involve people from diverse backgrounds during user testing of prototypes to check for issues related to representation and accessibility?

How do you evaluate whether the game prototype meets the intended educational objectives equitably across different demographics?

**Development**

Are accessibility features (e.g., subtitles, alternative control schemes) implemented at this stage to ensure the game is inclusive?

Are educational elements, such as quizzes or feedback loops, designed to be understandable and relevant for players from diverse backgrounds?

**Testing**

Do you specifically test for the effectiveness of representation and educational elements among underrepresented groups?

Do you include features? How do you ensure that accessibility features are working as intended during the testing phase?

**Deployment and Distribution**

Are steps taken to ensure that the game is accessible to low-income communities or regions with limited internet connectivity? Could you describe the steps?

Do you have a plan for collecting feedback on representation and educational content once the game is in the hands of the public?

**Maintenance and Updates**

Do you prioritize updates that may improve representation or accessibility? How?

Are there systems in place to update the educational content based on evolving best practices?

Are there systems feedback regarding its effectiveness across diverse groups?

**Marketing**

How do you balance the need balance equity features (representation, inclusivity,.etc ) with sales and marketing?

**Appendix B**

**Interview Notes**

Professional Background

- The interviewee is a Director of Software Engineering based in Kansas City.
- He has around 15 years of experience.

Game Development

- Although he has released some games, their expertise lies in AR (Augmented Reality) and general software development.
- He has worked on educational apps.

Representation and Diversity

- The interviewee acknowledges that the gaming industry is dominated by white men.
- He believes in involving people from diverse backgrounds in user testing.

Accessibility

- Accessibility features like subtitles are decided upon in the planning stage but may be implemented later.
- There's a consideration for low-income communities through various distribution platforms but limited connectivity in rural areas remains a challenge.

User Feedback and Equity

- They take user feedback seriously and sometimes incentivize it through 10-dollar gift cards.
- The interviewee believes that with targeted ad campaigns, it's possible to balance equity and marketing needs.

Internet Connectivity

- The lack of internet in rural areas is viewed as a human rights issue. The interviewee suggests increasing internet coverage to make games more accessible there.

Company Updates

- Updates that improve representation or accessibility are not usually a priority in their experience.